\documentclass[useAMS,usenatbib]{mn2e}
\usepackage{graphics}

\title[A List of Bright Interferometric Calibrators measured
at the ESO VLTI]
{A List of Bright Interferometric Calibrators measured
at the ESO VLTI
\thanks{
Based on observations made with ESO telescopes at Paranal Observatory
}
}

\author[A. Richichi et al.]{
        A. Richichi$^{1}$
\thanks{E-mail: arichich@eso.org}
   and
       I. Percheron$^{1}$
and
       J. Davis$^{2}$
       \\
$^{1}$European Southern Observatory, Karl-Schwarzschildstr.
2, D-85748 Garching bei M\"unchen, Germany\\
$^{2}$Sydney Institute for Astronomy, School of Physics,
University of Sydney, NSW 2006, Australia
}

\begin{document}
\date{Accepted 2009 June 19.  Received 2009 June 18; in original form 2008 June 9}

\pagerange{\pageref{firstpage}--\pageref{lastpage}} \pubyear{2008}

\maketitle
\label{firstpage}

\begin{abstract}
In a previous publication
\citep{paperi} we described a program of observations of
candidate calibrator stars at the ESO Very Large Telescope
Interferometer (VLTI), and presented the main results from a
statistical point of view. In the present paper, we concentrate on
establishing a new homogeneous group of bright interferometric
calibrators, based entirely on publicly available K-band VLTI
observations carried out with the VINCI instrument up to July
2004.  For this, we have defined a number of selection criteria
for the quality and volume of the observations, and we have
accordingly selected a list of 17 primary and 47 secondary
calibrators. We have developed an approach to a robust global fit
for the angular diameters using the whole volume of
quality-controlled data, largely independent of a priori
assumptions.  Our results have been compared with direct
measurements, and indirect estimates based on spectrophotometric
methods, and general agreement is found within the combined
uncertainties.  The stars in our list cover the range K=$-2.9$
to $+3.0$\,mag in brightness, and 1.3 to 20.5 milliarcseconds in
uniform-disk diameter. The relative accuracy of the angular
diameter values is on average 0.4\%  and 2\% for the primary and
secondary calibrators respectively. Our calibrators are well suited for
interferometric observations in the near-infrared on baselines
between $\approx 20$\,m and $\approx 200$\,m, and their accuracy
is superior, at least for the primary calibrators, to
other similar catalogues. Therefore, the present list
of calibrators has the potential to lead to
significantly improved interferometric scientific results.
\end{abstract}

%\titlerunning{-------------------}
%\maketitle

\section{Introduction}\label{introduction}

Optical long-baseline interferometry (OLBI) is the most powerful technique
available to overcome the limitations in angular resolution
imposed by the atmosphere, and by the diffraction limit of even
the largest and most perfect single mirror telescopes. By
combining the light from two or more telescopes separated by
distances much larger than the size of their mirrors, OLBI can
achieve angular resolutions at the level of one milliarcsecond
(mas) or less at optical and near-IR wavelengths. 
For illustration, 80\,m will suffice to completely resolve
a 1 milliarcsecond star at H$_\alpha$, while
about 300\,m will be required in the K band.
The astrophysical advantages of OLBI, as well as its challenging
technical difficulties, are described in reviews such as those by
\citet{quirr01} and \citet{monnier}, and in
the large lists of references they contain.

We do not detail here the numerous technical problems related to
the difficulty of performing high precision measurements while
maintaining an equal optical path between various telescopes.
However,  we note that a major limitation of OLBI is the need to
compare each measurement of a given science target with a
similar measurement of a star with known characteristics of
angular diameter and spectrum, made as close as possible in time
and angular distance. This process of calibration is necessary to
correct for the reduction in fringe visibility resulting from
instrumental effects and from the wavefront distortions caused by
the turbulent atmosphere.  The calibration factor is often called
the transfer function, and can be expressed as

\begin{equation}
{\rm TF} = V^2 / V^2_{\rm t} \label{eq:tf}
\end{equation}
where 
$V^2$ represents the observed
visibility squared, and
$V^2_{\rm t}$ is the theoretical visibility squared expected for an
observation without degradation due to instrumental or atmospheric
effects for the given source, baseline, and wavelength of
observation. Note that here, and in the rest of the paper, 
we refer to visibility squared as the squared modulus of 
the complex visibility.

Without entering into the merits and difficulties of obtaining a
reliable determination of $V^2$,
 it is clear that the
factor TF, and hence the final interferometric measurement, can
only be as accurate as our knowledge of the theoretical visibility
$V^2_{\rm t}$. This is ultimately defined by the accuracy with which we
know the angular diameter of the calibrator star - as well as
other characteristics such as its spectrum, which affects the
effective wavelength of the observation, and its limb darkening.

There are effectively two approaches to the problem of determining
the angular diameters of calibrator stars. One can use a
combination of spectrophotometric measurements and theoretical
predictions to derive empirical estimates. This has been done in
great detail for example by \citet{Coh99}, while
other authors have subsequently added more refined criteria
specifically for interferometric observations
(\citealt{Bor02}, \citealt{Mer05}).
Recently, \citet{Bonneau} have further refined this approach
by including information from online databases.  Alternatively, one
can list all available measurements obtained by angular resolution
methods, such as done by \citet{CHARM2}.
Obviously, both approaches have their advantages and
disadvantages. What is used in practice, at several
interferometric facilities around the world, is often a mixture of
all available information. This can easily lead to biased results.
A bias in the calibrator diameter is automatically reflected in a
systematic error on all derived interferometric measurements,
including the final scientific results. 
An extended discussion
of calibration errors has been provided by
\citet{VBVB}.
We also mention that one could in principle always find
suitably unresolved stars to be used as calibrators (i.e.
$V$=1 for all practical purposes), but this
has the disadvantage that in general these stars would be considerably
fainter than the scientific target and thus cause a loss
of quality in the results.
In principle 
it is also possible to avoid the use of calibrators
altogether by observing the first minimum in the visibility
curve of the scientific target. This was the initial 
approach by Michelson, however it presents the
disadvantage that it can only be used for sources
which are completely resolved and it involves
a number of complications that make it not really
suitable for modern, high-precision measurements.
We note 
however that, when using baselines with a significant
E-W component, the variation
of the visibility with hour angle has the potential
to determine an angular diameter without reference
to an overall multiplicative factor. This can be
used in principle to remove
ambiguities in the simultaneous determination
of the angular diameters of several stars from the same
data set, as is the case in the present work.
%Besides, this approach presents the practical difficulty
%of depending on precise measurements  of very low
%visibility values, of being critically dependent on
%detailed computations of
%bandwidth effects, and of requiring an optimal choice
%of baseline lengths for each target star.

In this paper, we derive a new homogeneous
list of interferometric calibrators, based on observations
collected solely with the ESO Very Large Telescope Interferometer
(VLTI, \citealt{Gli03}).
This is among the first attempts of its kind.
\citet{VB08} have recently published an investigation of
archival observations from the PTI, also carried out with the goal
of establishing a list of calibrators.  However, their approach differs
from ours in that it is based primarily on the statistical properties of
the observed visibilities and is restricted to essentially unresolved
stars whereas the main goal of our work has been to provide a list
of consistent and accurate angular diameter measurements.  The
fact that the PTI and the VLTI are in different hemispheres
suggests that the lists of
calibrators will be to a large extent complementary.  At present
we do not have access to the PTI list to establish what is expected
to be a minimal overlap.
For our goal,
we have started from the large volume of VLTI
commissioning data collected with the VINCI instrument for
candidate calibrator stars (as described in 
Richichi \& Percheron
\citeyear{paperi}, Paper~I hereafter),
and we have sought a self-consistent solution. 
The benefits of this approach
are a) the use of a single instrument and facility, with a well
understood behaviour in terms of optical and mechanical
characteristics; b) the use of a single automated process for the
data reduction; and c) relative freedom from possible bias due
to poor or wrong input values for the candidate calibrators.

In Sect.~\ref{calib} we recall the nature of the
observations,
we provide details of the criteria used to
select primary and secondary calibrators,
and we discuss our assumptions on the TF stability. 
In Sect.~\ref{method} we
describe two different approaches to  the global data analysis,
while in Sect.~\ref{results} we discuss the results obtained for
both primary and secondary calibrators, also in the light of other
available estimates and direct measurements. We conclude by
discussing the merits of this list and provide advice for its use,
as well as outlining possible improvements.
%}

\section{Definitions for a homogeneous, self-contained
list of calibrators}\label{calib}
\subsection{Data sets and initial selection}\label{datasets}
The data used in this paper, as well as in Paper~I, are extracted
from the large volume of VLTI public releases 
of VINCI data.
We have chosen them because of their large number, of the
relative simplicity of the VINCI instrument
which 
was designed to 
combine two
telescopes at a time in the undispersed K-band (about
2 to 2.4\,$\mu$m FWHM),
and of the availability of 
an automated pipeline which produces science-grade
results (Ballester et al. \citeyear{Bal04},
Kervella et al. \citeyear{Ker04}).
We mention that we used the fiber-optics version of VINCI,
i.e. the most commonly used. Other versions using integrated
optics were also briefly used. VINCI is now decommissioned.

In the VLTI scheme, the data are divided into observation blocks
(OBs). Each OB with VINCI contains a number of scans of the source,
typically 250. As reported in Paper~I, a total of 18352
VINCI OBs have been made publicly
available up to July 2004, covering 688 nights and involving 11
different baseline configurations using both the fixed 8.2\,m unit
telescopes (UT) and the relocatable 40\,cm siderostats.
Thanks to the VLTI design, and to the adoption of
optimized choices of parameters in the data reduction pipeline, the
results from 
the two sets of telescops can be considered homogeneous
and we make no distinction between them.
From Paper~I it can be seen that the baselines span
the range 8-140\,m, and that they have a wide range of position angles
with some emphasis on E-W directions.

The quality estimation criteria and the corresponding
results have been described in Paper~I. In particular, 12066 OBs
on objects classified as candidate calibrators were deemed of
sufficient quality. The list of candidate calibrators covered in
Paper~I included 191 stars, and a corresponding catalogue is
available on-line.

In Paper~I we defined the term candidate calibrator as a source
satisfying a number of criteria concerning its photometric
stability, the absence of nearby companions, and a spectral type
not subject to either sudden or long-term variability (see also
\citealt{Per03}). 
For the present purpose, we have further restricted 
ourselves to those stars believed
to have compact atmospheres, to be circularly symmetric, and without
wavelength effects in the spectral bandpass of the VINCI
instrument other than those attributable to a simple black body
energy distribution.
We have explicitly excluded
from the input list of
191 stars of Paper~I the following:
supergiants,
Cepheids and Miras; stars which are
non-circularly symmetric such as rapid rotators; double stars
with companions bright and/or close enough to affect the
measurement of visibility squared; and variable stars with
brightness variations $\ga 0.03$\,mag in the V band.

\subsection{Definition of primary and secondary calibrators}\label{pri_sec}
Not all stars were observed systematically and with
comparable accuracy, and
we have chosen to apply further selection criteria
as detailed
in Table~\ref{table_sel}.  The criteria are linked with a logical
AND. We have defined a list of primary calibrators, for which we
have imposed stringent quality criteria and also a high number of
repeated observations in the same night and across the whole
period under consideration. In parallel, we have established a
list of secondary calibrators, where the quality criteria have
been relaxed to some extent and for which the frequency and
distribution of observations (both in terms of dates and
baselines) are less stringent.

%\begin{table}[h]
\begin{table}
%\centering
\caption{Selection criteria for primary and secondary calibrators
% Criteria enforced for the
% selected sample of calibrators.
}
\label{table_sel}
\begin{tabular}{lcc}
\hline
\hline
\multicolumn{1}{c}{Criterion} &  Primary & Secondary \\
\hline
OBs / Night & $\ge 3$ & $\ge 3$\\
Nights & $\ge 3$ & $\ge 2$\\
Total OBs & $\ge 9$ &$\ge 8$\\
Baseline Ratio & $\ge 1.5$ & $>1$\\
Minimum $V^{2}$ & 0.05 & 0.05\\
Maximum $V^{2}$ & 0.85 & 0.85\\
%\hline \\
$\sigma V^2 / V^2$ & $\le 5$\% &$\le 7.5$\% \\
$<$Seeing$>$ / Night & $\le 1\farcs2$ &$\le 1\farcs4$ \\
$\sigma$Seeing / Night & $\le 0\farcs5$ & $\le 0\farcs5$ \\
\hline
Number of Selected Stars & 18 & 47\\
Total nights & 51 & 121\\
Total OBs   & 717 & 942\\
\hline
\hline
\end{tabular}
\end{table}

The baseline ratio is the ratio between the maximum and minimum
projected baseline length with which observations were collected
for the same star, and has been introduced to ensure a sufficient
span in the sampling of the visibility, at least for the primary
calibrators. In turn this provides a critical check on the fit to
the assumed model and allows possible discrepancies
to be identified. Note that in some cases the constraint of 1.5
for the baseline ratio among the primary calibrators had to
be partly relaxed after a careful check, as explained in
Sect.~\ref{analysis_primary}.
Minimum and maximum values for the observed
visibility squared were imposed. The minimum value was chosen to
reject values that might be affected by unknown low level
systematic errors or by significant bandwidth smearing (see
Appendix~\ref{app}).  It also
implicitly rejects all measurements carried out beyond the first
visibility minimum, thus avoiding possible issues in the fitting
process (such as ambiguities in the location of the minimum and
effects of limb-darkening). High values of the visibility
are important to test the quality of the fit and the possible
presence of contributions from other sources in the beam, but
at the high end they too can be biased (\citet{Ker04}).
For this reason a maximum threshold for the visibility squared was
also applied, which in any case is sufficiently high to ensure
a careful check of the model.

Table~\ref{table_sel} also lists some minimum
requirements on the quality of the night. For this, we used as
criteria the fractional error of the visibilities squared, as well
as the average and the standard deviation of the seeing for each
night. The former is a compromise between enforcing some intrinsic
quality on each data point, and the need to preserve a 
statistically meaningful sample.
The latter are computed from the DIMM measurements
available from the ESO ambient condition database, of which there
are generally of the order of a few hundred per night. It should
be noted that the DIMM values are in the visible.
For a few nights
during which the database was not operative, the average of the
DIMM measurements recorded at the beginning and end of each VLTI
data file were used. In this case, the data available for each
night may be significantly restricted, typically to a few tens of
points. Finally, there have been nights in which neither approach
was available, and in this case the data were rejected.
We have not explicitly added criteria on airmass $z$,
because observations at the VLTI are normally constrained
in zenith distance:
over 75\% of the OBs in our sample
were observed with $z\le 1.3$, and 90\% with $z\le1.5$.

%{\it Should we add that no visibility points were considered
%beyond the first minimum? This is implicit in the $V^2<0.05$
%criterium but might be worth stressing.
%Besides, we applied some rejection of nights for which
%the average transfer function of the interferometer
%was below 20%, due to optics, MONA, etc. Mention?
%Finally, mention that we had at least 3 stars/night.
%}
The criteria listed in Table~\ref{table_sel} are to some extent
subjective, but they are based on extensive experience gained in
analysing the total pool of data with a range of combinations of
criteria. The set of values chosen represents the best compromise
to ensure the selection of a sizable sample of primary calibrators
based on observational data of unquestionable quality, and a
sample of secondary calibrators of high quality which is
sufficiently large to provide satisfactory sky coverage.

Following the applications of these constraints,
the number of %potential 
primary calibrators was reduced to 17 and
these are listed in Table~\ref{tab:calib} 
(one of the potential primary calibrators,
$\theta$~Cet, was removed for other reasons 
as discussed in Sect.~\ref{analysis_primary}).
%, reducing the number
%of primary calibrators to 17). 
In Table~\ref{tab:calib} we provide
the most common identifiers, the $V$ and $K$ magnitudes and the
spectral class.  An effective temperature has been adopted for
each star. Under the assumption of a black body,
the effective wavelength of the VINCI observation
for each OB has been computed and is also listed in
Table~\ref{tab:calib}. 
The computation was based on an
instrumental spectral response which includes the properties of
the filter, telescopes, interferometer optics, and detector
\citep{Davis03}.  Baseline uncertainties are estimated
to be negligible and have been ignored. The last four columns of
Table~\ref{tab:calib} report the total number of nights on which
each calibrator was observed and the total number of OBs
available. The columns marked as `Pre' list the numbers after the
pre-selection made by applying the criteria of
Table~\ref{table_sel}. However, additional criteria were 
subsequently applied, and the final numbers are listed in the
columns marked as `Fin'. This is discussed in Sect.~\ref{analysis_primary}.
With the same conventions and prescriptions,
the list of potential secondary calibrators is shown in
Table~\ref{table_sample2}.
%Table~\ref{tab:calib}.

\begin{table*}
%\centering
\begin{minipage}{140mm}
\caption{List of the %potential 
primary calibrators and their
observational statistics. 
%As explained in 
%Sect.~\ref{method},
%{$\theta$ Cet} was excluded from the final list.
}
\label{tab:calib}
\begin{tabular}{rrcrrlccccc}
\hline
\hline
HR & \multicolumn{1}{c}{HD} & Star & \multicolumn{1}{c}{V} &
\multicolumn{1}{c}{K} & \multicolumn{1}{c}{Spectral} &
$\lambda_{\mathrm{eff}}$ & \multicolumn{2}{c}{Nights} & \multicolumn{2}{c}{OBs} \\
   &    &      &   &   & \multicolumn{1}{c}{Class} &  ($\mu$m) & Pre & Fin
& Pre & Fin \\
\hline

        74 &       1522 &  $\iota$ Cet  &       3.56 &       1.03 &  K1.5 III  &      2.184 &          9 &          8 &         54 &         48 \\

       188 &       4128 &  $\beta$ Cet  &       2.04 &      -0.27 &    K0 III  &      2.184 &          8 &          7 &         41 &         38 \\

%       402 &       8512 &  $\theta$ Cet  &       2.06 &      -0.35 &   K0 IIIb %&      2.184 &          3 &          0 &         10 &          0 \\
%
       911 &      18884 &  $\alpha$ Cet  &       2.56 &      -1.68 &  M1.5 IIIa  &      2.185 &          4 &          4 &         15 &         14 \\

      1231 &      25025 &  $\gamma$ Eri  &       2.98 &      -0.95 &   M1 IIIb  &      2.185 &          4 &          4 &         22 &         22 \\

      2491 &      48915 &  $\alpha$ CMa  &      -1.47 &      -1.39 &      A1 V  &      2.181 &          7 &          6 &         29 &         26 \\

      2943 &      61421 &  $\alpha$ CMi A  &       0.34 &      -0.55 &   F5 IV-V  &      2.182 &          3 &          3 &         20 &         20 \\

      3748 &      81797 &  $\alpha$ Hya  &       2.00 &      -1.13 &  K3 II-III  &      2.184 &          6 &          6 &         59 &         56 \\

      4050 &      89388 &  V337 Car  &       3.38 &      -0.09 &    K3 IIa  &      2.184 &          3 &          3 &         15 &         15 \\

      4232 &      93813 &  $\nu$ Hya  &       3.11 &       0.25 &  K0/K1 III  &      2.184 &          4 &          4 &         18 &         18 \\

      4546 &     102964 &   HR 4546  &       4.47 &       1.52 &    K3 III  &      2.184 &          5 &          5 &         24 &         24 \\

      4630 &     105707 &  $\epsilon$ Crv  &       3.02 &      -0.06 &    K2 III  &      2.184 &         10 &         10 &         36 &         36 \\

      5288 &     123139 &  $\theta$ Cen  &       2.06 &      -0.35 &   K0 IIIb  &      2.184 &         21 &         21 &        136 &        136 \\

      6048 &     145897 &  $\chi$ Sco  &       5.25 &       2.28 &    K3 III  &      2.184 &          5 &          5 &         18 &         18 \\

      6056 &     146051 &  $\delta$ Oph  &       2.74 &      -1.18 &  M0.5 III  &      2.185 &          6 &          6 &         93 &         93 \\

      6241 &     151680 &  $\epsilon$ Sco  &       2.29 &      -0.39 &  K2.5 III  &      2.184 &          8 &          8 &         38 &         37 \\

      7873 &     196321 &    70 Aql  &       4.90 &       1.10 &     K5 II  &      2.185 &         12 &         11 &         57 &         54 \\

      8411 &     209688 &  $\lambda$ Gru  &       4.48 &       1.37 &    K3 III  &      2.184 &          7 &          6 &         33 &         29 \\

\hline \hline
\end{tabular}
\end{minipage}
\end{table*}

\subsection{Stability of the transfer function}\label{tra_fun}
A crucial assumption in our approach is that
the transfer
function remained constant throughout each night, i.e. one unique
value for each night. 
A correlation between
seeing and transfer function is commonly accepted but quantitative
details remain largely unstudied. Work by
\citet{Davis05} has shown that the assumption of a transfer
function constant throughout a whole night can be a good
approximation at the Paranal site, but this was by no means a
comprehensive study. Results by  \citet{Leb06}
have pointed to variability of the transfer function also
in Paranal, although under less than ideal seeing conditions.
Interestingly, a separate recent study by \citet{Leb09} 
has shown a very stable transfer function.
The analysis of a few selected nights by \citet{iza07} showed that
the stability of the transfer function is not uncommon, but is
not guaranteed either.
There is no easy solution to this problem, also considering
that in general the VLTI calibrator observations for any given night
are interleaved with gaps in time which can amount to many hours,
and it is necessary to find a viable compromise.
By imposing criteria on the average
seeing and its variation, we have sought to restrict our sample to
nights with good atmospheric conditions. 
Other parameters which have been shown to correlate well
with the transfer function are atmospheric jitter
\citep{Col99}, which unfortunately 
is not readily available for our sample, and coherence time.
We have plotted coherence time against seeing for almost
the totality of the nights under consideration, and
we have found  a marked correlation, with only about 4\% of
the nights showing a long coherence time when the
seeing was relatively poor (i.e. seeing seems to be
a tighter contraint than coherence time). It should be
noted that both these parameters are estimated at
visible wavelengths,
for a star other than the one observed
at the VLTI,
and at a location on the Paranal
summit which is not coincident with the interferometer.
Recently the VLTI has been complemented with
a fringe-tracker (FINITO, Corcione \citeyear{Corc03}) which is
already providing accurate results 
(Spaleniak \citeyear{iza07},
Le~Bouquin \citeyear{Leb08}) 
and will progressively alleviate some of these problems.

\begin{table*}
%\centering
%\centering
\begin{minipage}{140mm}
\caption{Same as Table~\ref{tab:calib},
for the secondary calibrators.
}
\label{table_sample2}
\begin{tabular}{rrcrrlccccc}
\hline
\hline
HR & \multicolumn{1}{c}{HD} & Star & \multicolumn{1}{c}{V} &
\multicolumn{1}{c}{K} & \multicolumn{1}{c}{Spectral} &
$\lambda_{\mathrm{eff}}$ & \multicolumn{2}{c}{Nights} & \multicolumn{2}{c}{OBs} \\
   &    &      &   &   & \multicolumn{1}{c}{Class} &  ($\mu$m) & Pre & Fin
& Pre & Fin \\
\hline

        37 &        787 &      HR 37 &       5.29 &       1.86 &      K4III &      2.184 &          2 &          2 &          9 &          8 \\

       248 &       5112 &     20 Cet &       4.78 &       1.02 &      M0III &      2.185 &          7 &          7 &         38 &         38 \\

       402 &       8512 & $\theta$ Cet &       3.60 &       1.29 &      K0III &      2.184 &         10 &          8 &         32 &         26 \\

       440 &       9362 & $\delta$ Phe &       3.95 &       1.63 &      G9III &      2.184 &          4 &          1 &         14 &          3 \\

       509 &      10700 & $\tau$ Cet &       3.50 &       1.79 &        G8V &      2.184 &          6 &          3 &         20 &         11 \\

       602 &      12524 & $\chi$ Phe &       5.16 &       1.57 &      K5III &      2.185 &          6 &          3 &         39 &         17 \\

      1084 &      22049 & $\epsilon$ Eri &       3.73 &       1.78 &        K2V &      2.184 &          6 &          6 &         23 &         23 \\

      1136 &      23249 & $\delta$ Eri &       3.51 &       1.62 &       K0IV &      2.184 &          4 &          4 &         17 &         16 \\

      1318 &      26846 &   39 Eri A &       4.90 &       2.26 &      K3III &      2.184 &          2 &          2 &         21 &         21 \\

      1652 &      32831 & $\gamma_1$ Cae &       4.58 &       1.76 &      K3III &      2.184 &          8 &          7 &         38 &         34 \\

      1663 &      33042 & $\eta_2$ Pic &       5.06 &       1.52 &      K5III &      2.185 &          7 &          6 &         34 &         31 \\

      1713 &      34085 & $\beta$ Ori &       0.12 &       0.21 &     B8Iab: &      2.180 &          7 &          5 &         29 &         16 \\

      1799 &      35536 &    HR 1799 &       5.61 &       2.13 &      K5III &      2.185 &          4 &          4 &         18 &         18 \\

      1829 &      36079 & $\beta$ Lep &       2.84 &       0.95 &       G5II &      2.183 &          5 &          5 &         37 &         37 \\

      1834 &      36167 &     31 Ori &       4.71 &       0.81 &      K5III &      2.185 &          3 &          3 &         15 &         15 \\

      2311 &      45018 &    HR 2311 &       5.62 &       1.70 &      K5III &      2.185 &          2 &          1 &          8 &          3 \\

      2478 &      48433 &     30 Gem &       4.50 &       1.86 &      K1III &      2.184 &          2 &          2 &          8 &          8 \\

      2503 &      49161 &     17 Mon &       4.78 &       1.58 &      K4III &      2.184 &          4 &          4 &         19 &         19 \\

      2574 &      50778 & $\theta$ CMa &       4.09 &       0.64 &      K4III &      2.184 &          4 &          4 &         14 &         14 \\

      2693 &      54605 & $\delta$ CMa &       1.84 &       0.45 &     F8Iab: &      2.182 &          2 &          1 &         10 &          4 \\

      2736 &      55865 & $\gamma_2$ Vol &       3.77 &       1.44 &      K0III &      2.184 &          3 &          2 &         19 &         13 \\

      3046 &      63744 &    HR 3046 &       4.70 &       2.31 &      K0III &      2.184 &          3 &          3 &         12 &         12 \\

      3845 &      83618 & $\iota$ Hya &       3.91 &       0.87 &    K2.5III &      2.184 &          7 &          3 &         29 &         13 \\

      3980 &      87837 &     31 Leo &       4.38 &       0.98 &      K4III &      2.184 &          4 &          3 &         20 &         15 \\

      4450 &     100407 &  $\xi$ Hya &       3.54 &       1.47 &      G7III &      2.183 &          3 &          2 &         15 &          9 \\

      4526 &     102461 &   V918 Cen &       5.44 &       1.38 &      K5III &      2.185 &          3 &          2 &         11 &          6 \\

      4831 &     110458 &    HR 4831 &       4.67 &       2.23 &      K0III &      2.184 &          5 &          2 &         21 &          7 \\

      5287 &     123123 &  $\pi$ Hya &       3.26 &       0.75 &      K2III &      2.184 &          3 &          2 &          9 &          6 \\

      5340 &     124897 & $\alpha$ Boo &      -0.04 &      -2.91 &    K1.5III &      2.184 &          4 &          1 &         18 &          3 \\

      5381 &     125932 &     51 Hya &       4.77 &       1.77 &      K5III &      2.185 &          2 &          2 &          9 &          9 \\

      5487 &     129502 &  $\mu$ Vir &       3.90 &       3.04 &      F2III &      2.181 &          2 &          2 &          9 &          9 \\

      5513 &     130157 &    HR 5513 &       6.07 &       2.13 &   K4/K5III &      2.185 &          4 &          2 &         18 &          9 \\

      5526 &     130694 &     58 Hya &       4.42 &       1.11 &      K4III &      2.184 &          4 &          3 &         15 &         12 \\

      5705 &     136422 & $\phi_1$ Lup &       3.58 &      -0.15 &      K5III &      2.185 &          5 &          1 &         23 &          6 \\

      6229 &     151249 & $\eta$ Ara &       3.78 &      -0.12 &      K5III &      2.185 &          2 &          2 &          9 &          9 \\

      6630 &     161892 &    HR 6630 &       3.19 &       0.62 &      K2III &      2.184 &          2 &          2 &         10 &         10 \\

      6862 &     168592 &    HR 6862 &       5.09 &       1.57 &    K4.5III &      2.185 &          2 &          1 &          8 &          3 \\

      6913 &     169916 & $\lambda$ Sgr &       2.83 &       0.40 &     K1IIIb &      2.184 &          6 &          5 &         26 &         21 \\

      7234 &     177716 & $\tau$ Sgr &       3.32 &       0.59 &     K1IIIb &      2.184 &          2 &          1 &          9 &          6 \\

      7557 &     187642 & $\alpha$ Aql &       0.77 &       0.07 &        A7V &      2.181 &          4 &          3 &         15 &         12 \\

      7584 &     188154 &     56 Aql &       5.79 &       1.70 &      K5III &      2.185 &          7 &          5 &         27 &         21 \\

      7635 &     189319 & $\gamma$ Sge &       3.53 &      -0.23 &      M0III &      2.185 &          3 &          3 &         12 &         12 \\

      7869 &     196171 & $\alpha$ Ind &       3.12 &       0.86 & K0IIICNvar... &      2.184 &          6 &          3 &         30 &         12 \\

      8080 &     200914 &     24 Cap &       4.53 &       0.54 &   K5/M0III &      2.185 &         14 &         14 &         67 &         67 \\

      8679 &     216032 & $\tau$ Aqr &       4.05 &       0.22 &      K5III &      2.185 &          3 &          2 &         20 &         14 \\

      8685 &     216149 &    HR 8685 &       5.41 &       2.10 &      M0III &      2.185 &          5 &          4 &         28 &         22 \\

      8812 &     218594 &     88 Aqr &       3.66 &       0.94 &      K1III &      2.184 &          3 &          3 &         10 &         10 \\

\hline
\hline
\end{tabular}
\end{minipage}
\end{table*}

\section{The problem and its solutions}\label{method}
Two approaches have been applied to the problem of extracting
consistent angular diameters from the available set of observed
calibrator data: a) the analysis of the data on a night-by-night
basis (NN), and b) a global solution of the entire data set (GS).

\begin{figure}
%\vspace{8cm}
\resizebox{\hsize}{!}{\includegraphics{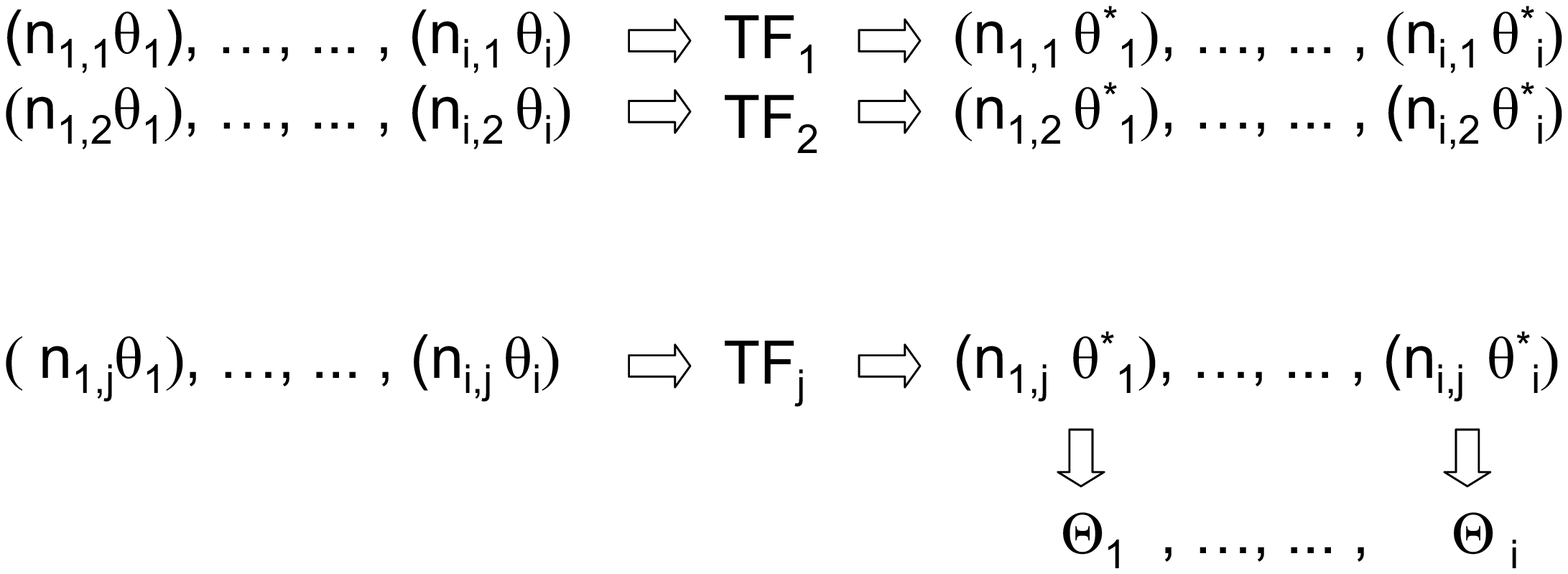}}
\caption{
Schematic description of the night-by-night (NN) approach to the
determination of revised diameters $\Theta$ of $i$ calibrators
observed over $j$ nights.  The symbol n$_{\rm{i,j}}$ denotes the
number of measurements available per each calibrator and night.
The left-to-right arrows indicate processing done independently
for each night. The top-to-bottom arrows indicate an averaging
process done across all the nights.}\label{fig:nn}
\end{figure}

In the NN approach, for each night, starting from a set of initial
guesses for the angular diameters, the best value of the transfer
function that minimises the spread of calibrator observations is
determined. This transfer function is then used to determine new
calibrator diameters independently of possible values from other
nights. Effectively, this corresponds to solving a relatively
large number (equal to the number of available nights) of small
independent problems. Each of these has few unknowns, namely the
angular diameters of the calibrators observed on the given night.
The number of unknowns varied from a minimum of three to a maximum
of six for the primary calibrators. At the end of the process, 
several independent
angular diameter determinations  are available for each calibrator,
and the average is computed across all the nights.
A schematic description is
given in Fig.~\ref{fig:nn}, where we have denoted with
$\theta_{\rm i}$ the initial angular diameters of the calibrators,
and with $\theta^{\rm *}_{\rm i}$ the corrected values.

\begin{figure}
%\vspace{8cm}
\resizebox{\hsize}{!}{\includegraphics{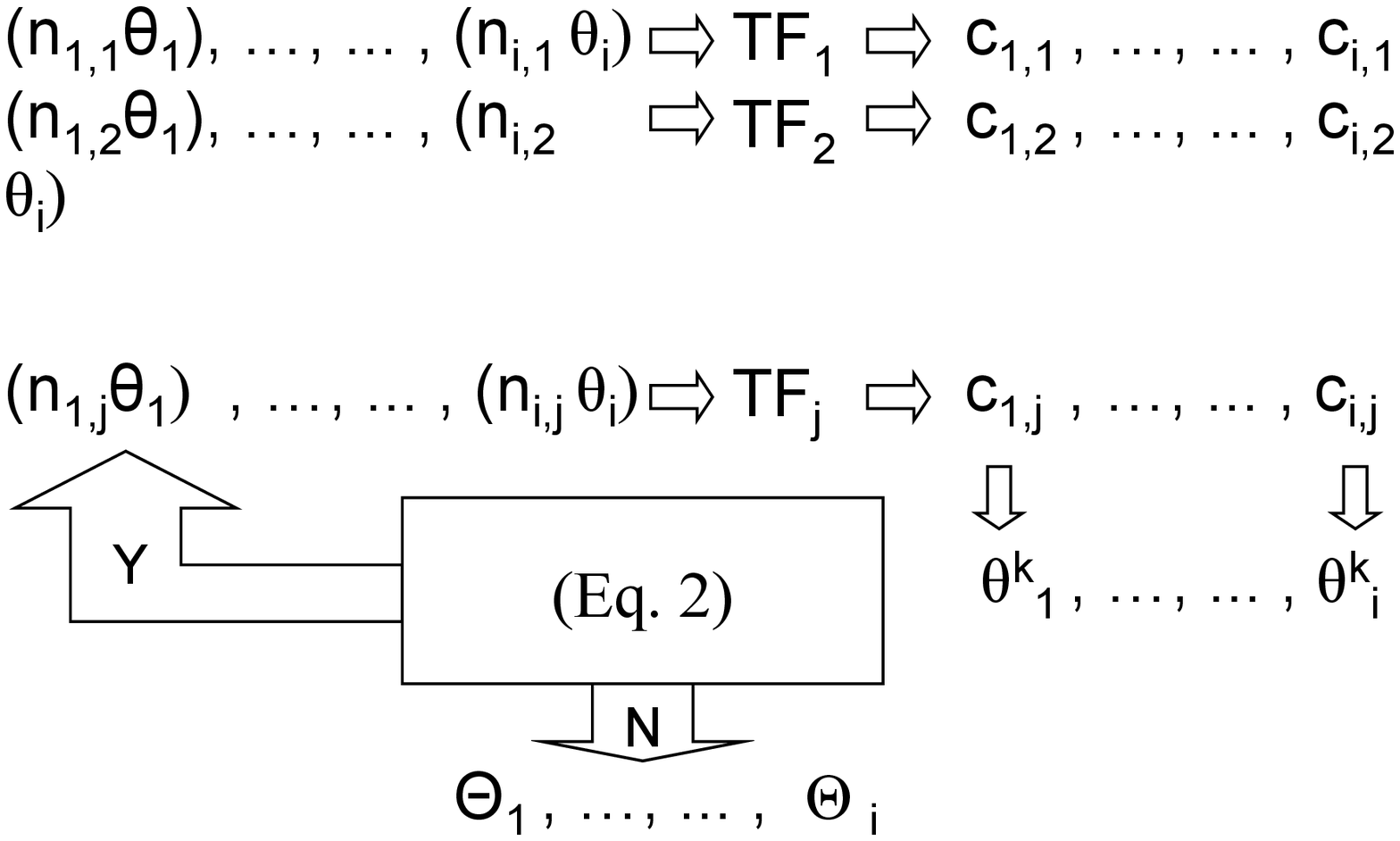}}
\caption[ ]{
Same as Fig.~\ref{fig:nn}, for the
global solution (GS)  approach.
Here, the c$_{\rm{i,j}}$ symbols denote the
calibrated visibilities squared, and the top-down arrows
represent a global fit for all available
nights and baselines.
The rectangle represents the decision criterion to continue
or stop iterations.
}\label{fig:gs}
\end{figure}

The GS approach, on the other hand, involves the solution  of a
single problem with many more unknowns (all the individual angular
diameters) and equations (all available nights). The starting
point is similar to the NN approach, i.e. a set of initial values
for the angular diameters is adopted and a best-fitting transfer
function is determined for each night. However, instead of
determining the corresponding revised angular diameters for each
night, we retain the calibrated values of visibility squared and
merge them for all nights for each calibrator.  The angular
diameter fit is then made to the merged values of visibility
squared to obtain a global value for the angular diameter.  
The process is then iterated starting with the new set of angular
diameters.  This is represented 
schematically in Fig.~\ref{fig:gs}.  

Convergence is examined independently for each calibrator by
comparing the values of its angular diameter $\theta_{\rm i}^{k}$
and error $\Delta \theta_{\rm i}^{k}$ determined at each global
iteration $k$ with the previous one. Changes in the angular
diameter are applied until both of the following conditions are
satisfied:

\begin{equation}
|\theta_{\rm i}^{k}-\theta_{\rm i}^{k-1}| \le \beta
\Delta\theta_{\rm i}^{k}
\\
{\chi^2}_{\rm i}^{k} < {\chi^2}_{\rm i}^{k-1}
\label{eq:tfgs}
\end{equation}
where $\beta$ is a factor between zero and unity and $\chi^2$
is computed from the fitted model and the input calibrated
visibilities squared and errors for each calibrator. It should be noted
that the angular diameters for all calibrators are free at each
iteration, and the above constraints only determine when they are
actually changed.  After some trials, we found that $\beta=0.5$
was a good compromise ensuring sufficient freedom for variation
and, at the same time, limiting the iterations to a reasonable
number ($<20$). 
Global convergence was considered to have been achieved
when no further change to any angular diameter was required according
to Eq.~\ref{eq:tfgs}.

%The GS  approach is more
%vulnerable to the details of the input data.
%For example,
%calibrated visibilities squared 
%from different nights might have the same
%formal accuracy but, if they refer to different baselines, they can
%have a very different influence on the diameter fit. Besides, 
%the problem has been proven to be ill-posed in the case
%of N-S baselines (referee communication). Our case is different
%since we have a wide range of baselines including many generally
%oriented E-W. This adds a considerable additional information,
%since each night provides in practice several (projected) baselines
%when the stars are observed over a significant amount of time.
%Nonetheless we
%were concerned with possible convergence issues depending on the
%quality of the initial guesses for the angular diameters.
%This has been tested and accounted for with an additional
%layer of MonteCarlo simulations, as described in Sect.~\ref{analysis_sec}.

\subsection{Analysis of the primary calibrators}\label{analysis_primary}
For the primary calibrators
we implemented first the NN method, which
enables discrepant values to be identified on a nightly basis.
Thus a very good set of initial angular diameter
estimates could be established
as a starting point for the GS method and also as a benchmark for comparisons.

Initial best estimates for the uniform-disk
angular diameters were adopted from Paper~I and were employed to
compute values for the transfer function (TF), as given by
Eq.~\ref{eq:tf}, for each OB. The mean TF for the night was
computed and used to determine a revised uniform-disk angular
diameter for each star  on each night. 
\begin{figure}
%\vspace{8cm}
\resizebox{\hsize}{!}{\includegraphics{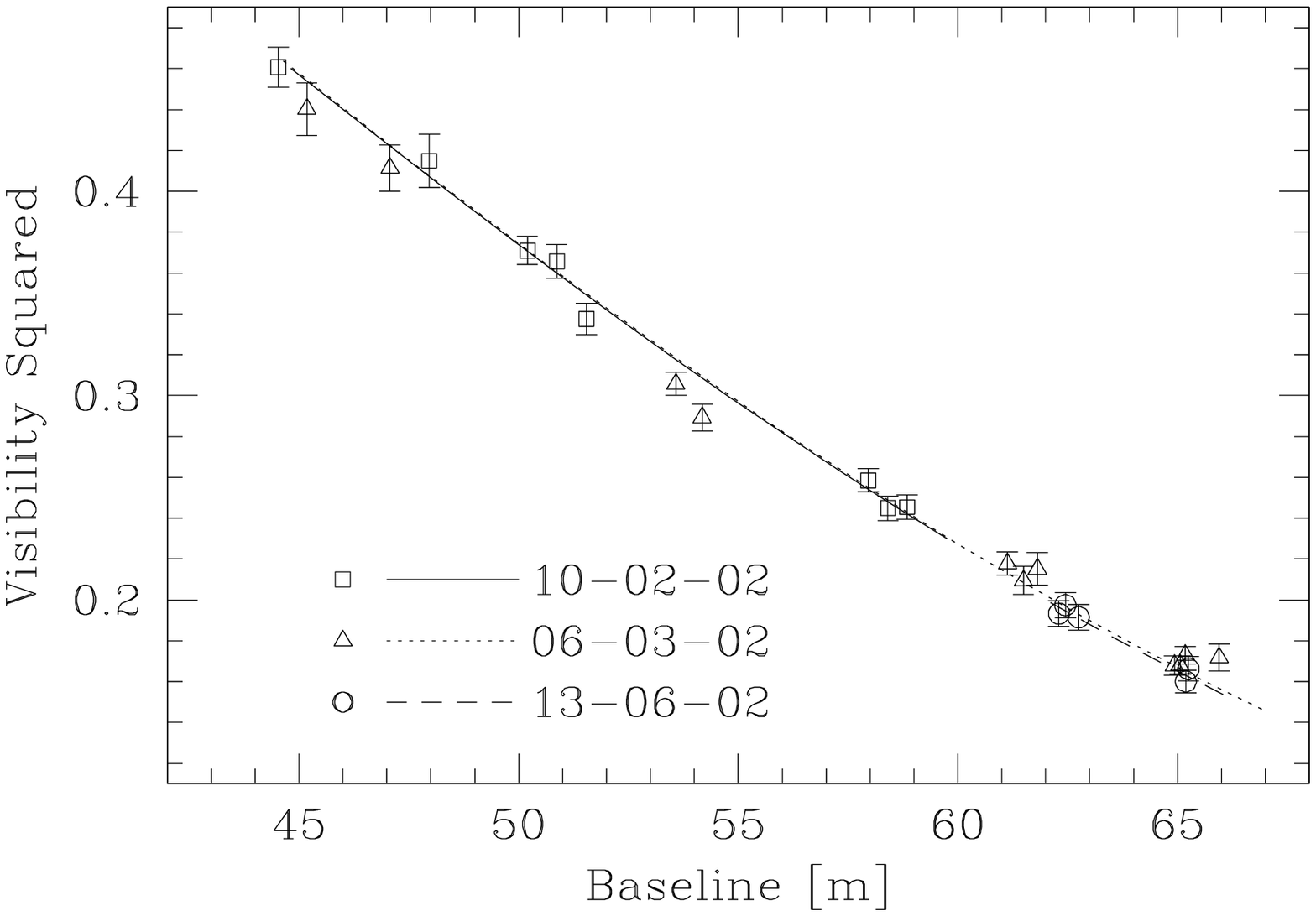}}
\caption[ ]{
Three data sets of the star $\theta$~Cen,
out of a total of twenty-one nights available, analyzed
in the night-by-night (NN) approach.
The symbols and lines for the three dates shown correspond
to calibrated visibilities squared and fits. The fits
are for the UD values of
$5.451\pm0.020$,
$5.446\pm0.015$ and
$5.461\pm0.023$ mas, in the order shown.
The reduced $\chi^2$ values
are
0.567,
2.498 and
0.238,
respectively.
The lines are almost overlapping.
}\label{fig:tetcen1}
\end{figure}

In principle, the process can be repeated, but we found that the
diameters did not change after the first iteration and that the
process is insensitive to the choice of initial values. The number
of observations $n$ of each star on each night is either zero, or
$\ge 3$ (see Table~\ref{table_sel}).
Fig.~\ref{fig:tetcen1} illustrates one example of
the breakdown of the results
on a nightly basis intrinsic in the NN approach.
After all the nights had
been processed, final estimates for the angular diameters
$\Theta_{\rm i}$ were computed.
We opted to use 
simple means and standard errors,
rather than weighted means and errors. This is justified by
the intrinsic spread in the data, and by the fact that these
values are only meant to serve as input for the GS approach. In
Table~\ref{tab:nn} we list the resulting uniform-disk angular
diameters, the limb-darkening factors as discussed in Appendix~\ref{app},
and the derived limb-darkened angular diameters.

%It should be noted that 
%at this point the list was reduced from 18
%(see Table~\ref{tab:calib}) to 17 stars, because 
Further data selection
was also applied at this stage. 
In particular, we identified a small number of
discrepant points for some stars and removed them: this was
justified when the measured value lay several standard deviations
away from the other values for the same star in the same night,
and plausible instrumental or observational reasons could be
identified. We also required that each night should have at least
three primary calibrators observed in order to derive a robust
estimate of the transfer function. Finally, we took into
consideration the fact that the VLTI transfer function was
sometimes low due to understood technical reasons (see Paper~I for
a discussion).  In principle the data for these nights are
consistent but, for added confidence in the results, we excluded
nights in which the transfer function was below 0.2. As  a result,
some OBs and even some nights had to be removed, and one star was
dropped from our final list ($\theta$~Cet, which was subsequently
selected as a secondary calibrator).  Finally, we note that as a
side-effect of this additional filtering, the constraint on the
baseline ratio (see Table~\ref{table_sel}) could not be preserved
in all cases. In particular, for about half of the stars the
effective span of baselines covers a ratio of 1.2-1.3 rather than
the 1.5 initially specified.  For these stars we satisfied
ourselves, by inspecting the distribution of visibilities, that
the model fit was constrained convincingly and we have retained
them.

\begin{table}
%\centering
\caption{List of the
angular diameters for the
primary calibrators,
based on the night-by-night (NN) approach.
}
\label{tab:nn}
\begin{tabular}{lrcr}
\hline
\hline
Star & \multicolumn{1}{c}{$\theta_{\mathrm{UD}}$(mas)} &
$\rho_{\lambda}$ & \multicolumn{1}{c}{$\theta_{\mathrm{LD}}$(mas)} \\
\hline
$\iota$  Cet     & $ 3.228 \pm 0.046 $ & 1.0245 & $ 3.307 \pm 0.047 $ \\
$\beta$  Cet     & $ 5.191 \pm 0.008 $ & 1.0228 & $ 5.309 \pm 0.008 $ \\
$\alpha$ Cet     & $ 12.007 \pm 0.295 $ & 1.0297 & $ 12.364 \pm 0.304 $ \\
$\gamma$ Eri     & $ 8.654 \pm 0.448 $ & 1.0293 & $ 8.908 \pm 0.461 $ \\
$\alpha$ CMa     & $ 5.953 \pm 0.082 $ & 1.0097 & $ 6.011 \pm 0.083 $ \\
$\alpha$ CMi A    & $ 5.288 \pm 0.077 $ & 1.0151 & $ 5.368 \pm 0.078 $ \\
$\alpha$ Hya    & $ 9.384 \pm 0.242 $ & 1.0258 & $ 9.626 \pm 0.248 $ \\
V337 Car     & $ 5.245 \pm 0.046 $ & 1.0264 & $ 5.383 \pm 0.047 $ \\
$\nu$ Hya    & $ 4.489 \pm 0.012 $ & 1.0254 & $ 4.603 \pm 0.012 $ \\
HR 4546      & $ 2.473 \pm 0.011 $ & 1.0258 & $ 2.537 \pm 0.011 $ \\
$\epsilon$ Crv   & $ 4.901 \pm 0.045 $ & 1.0254 & $ 5.025 \pm 0.046 $ \\
$\theta$ Cen   & $ 5.412 \pm 0.066 $ & 1.0229 & $ 5.536 \pm 0.068 $ \\
$\chi$ Sco   & $ 2.117 \pm 0.031 $ & 1.0255 & $ 2.171 \pm 0.032 $ \\
$\delta$ Oph   & $ 9.642 \pm 0.067 $ & 1.0293 & $ 9.925 \pm 0.069 $ \\
$\epsilon$ Sco   & $ 5.640 \pm 0.019 $ & 1.0239 & $ 5.775 \pm 0.019 $ \\
70 Aql   & $ 3.259 \pm 0.094 $ & 1.0281 & $ 3.351 \pm 0.097 $ \\
$\lambda$ Gru   & $ 2.619 \pm 0.094 $ & 1.0258 & $ 2.687 \pm 0.096 $ \\
\hline \hline
\end{tabular}
\end{table}

\begin{figure}
%\vspace{8cm}
\resizebox{\hsize}{!}{\includegraphics{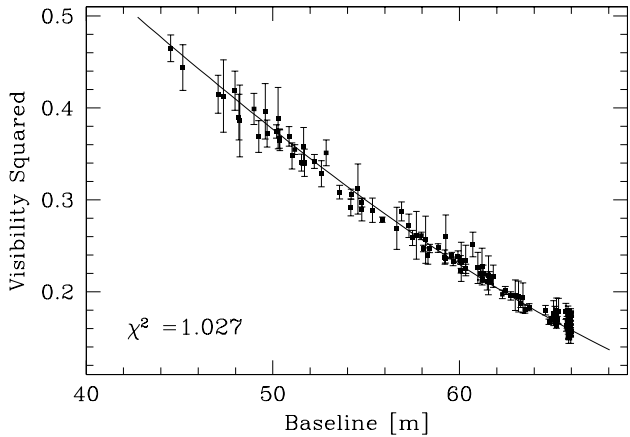}}
\caption[ ]{
The data set of the star $\theta$~Cen,
for the total of 21 nights available,
analyzed
in the global solution (GS) approach.
The solid line is the fit for a uniform-disk
diameter of $5.431$\,mas.
The normalized $\chi^2$ is indicated.
}\label{fig:tetcen2}
\end{figure}

Subsequently, we applied the GS algorithm, using the NN values
given in Table~\ref{tab:nn} as a starting point.
Fig.~\ref{fig:tetcen2} shows one example of how the GS
approach extends over all the available data and nights,
and should be compared with
Fig.~\ref{fig:tetcen1}.
The resulting uniform-disk angular diameter values are
listed in Table~\ref{tab:gs} together with the differences between
the GS and NN values. The limb-darkened angular diameters,
obtained using the $\rho_\lambda$ coefficients listed in
Table~\ref{tab:nn}, are also listed in Table~\ref{tab:gs}.

\begin{table}
%\centering
\caption{List of the
angular diameters for the
primary calibrators,
based on the global solution (GS) approach.
The error for 
%{$\epsilon$ Crv} 
{$\alpha$ CMa}
has
been doubled, to account for bandwidth smearing
which is described in the Appendix.
}
\label{tab:gs}
\begin{tabular}{crrc}
\hline
\hline
Star &
\multicolumn{1}{c}{$\theta_{\mathrm{UD}}$(mas)} &
\multicolumn{1}{c}{$\theta_{\mathrm{LD}}$(mas)} &
\multicolumn{1}{c}{$\theta_{\mathrm{UD}}^{\mathrm{GS}}$-
                   $\theta_{\mathrm{UD}}^{\mathrm{NN}}$} \\
\hline
$\iota$  Cet     & $ 3.245 \pm 0.010 $ & $ 3.325 \pm 0.010 $ & $+$0.017 \\
$\beta$  Cet     & $ 5.210 \pm 0.005 $ & $ 5.329 \pm 0.005 $ & $+$0.019 \\
$\alpha$ Cet     & $ 12.216 \pm 0.075 $ & $ 12.579 \pm 0.078 $ & $+$0.209 \\
$\gamma$ Eri     & $ 8.389 \pm 0.076 $ & $ 8.634 \pm 0.078 $ & $-$0.265 \\
%$\alpha$ CMa     & $ 6.030 \pm {\bf 0.020} $ & $ 6.089 \pm {\bf 0.020} $ & $+$0.077 \\
$\alpha$ CMa     & $ 6.030 \pm { 0.020} $ & $ 6.089 \pm { 0.020} $ & $+$0.077 \\
$\alpha$ CMi A    & $ 5.357 \pm 0.023 $ & $ 5.438 \pm 0.023 $ & $+$0.069 \\
$\alpha$ Hya    & $ 9.100 \pm 0.016 $ & $ 9.335 \pm 0.016 $ & $-$0.284 \\
V337 Car     & $ 5.270 \pm 0.024 $ & $ 5.409 \pm 0.025 $ & $+$0.025 \\
$\nu$ Hya    & $ 4.537 \pm 0.014 $ & $ 4.652 \pm 0.015 $ & $+$0.048 \\
HR 4546      & $ 2.475 \pm 0.004 $ & $ 2.538 \pm 0.004 $ & $+$0.002 \\
%$\epsilon$ Crv   & $ 4.897 \pm {\bf 0.009} $ & $ 5.021 \pm {\bf 0.010} $ & $-$0.005 \\
$\epsilon$ Crv   & $ 4.897 \pm { 0.009} $ & $ 5.021 \pm { 0.010} $ & $-$0.005 \\
$\theta$ Cen   & $ 5.431 \pm 0.004 $ & $ 5.556 \pm 0.005 $ & $+$0.019 \\
$\chi$ Sco   & $ 2.121 \pm 0.013 $ & $ 2.176 \pm 0.013 $ & $+$0.004 \\
$\delta$ Oph   & $ 9.663 \pm 0.013 $ & $ 9.946 \pm 0.013 $ & $+$0.021 \\
$\epsilon$ Sco   & $ 5.612 \pm 0.008 $ & $ 5.747 \pm 0.008 $ & $-$0.028 \\
70 Aql   & $ 3.246 \pm 0.014 $ & $ 3.337 \pm 0.014 $ & $-$0.013 \\
$\lambda$ Gru   & $ 2.635 \pm 0.020 $ & $ 2.703 \pm 0.020 $ & $+$0.016 \\

\hline \hline
\end{tabular}
\end{table}

The differences between the NN and GS results are small, both in
absolute and relative terms, confirming that the input values from
the NN approach were already quite accurate. The average of the
last column in Table~\ref{tab:gs} is 4 microarseconds, indicating
that there is no systematic bias towards larger or smaller angular
diameters between the two approaches.

\subsection{Analysis of the secondary calibrators}\label{analysis_sec}
For the secondary calibrators (SCs hereafter), the GS approach was
employed directly; the data set consisted of one squared
visibility amplitude for every OB, and were fit with
one stellar diameter for each SC and one TF for each night.
In applying the GS approach to the SCs,
we used both the primary and secondary calibrators available on
each night. 
The PC diameters were fixed with the values
of Table~\ref{tab:gs}, and only the SC diameters were free to vary
according to the criteria of Eq.~\ref{eq:tfgs}.  
While for the primary calibrators (PCs hereafter) we
used the NN approach to obtain a set of good starting values for
the angular diameters, for the SCs this was not always
possible. In addition to a relatively more relaxed set of
constraints (see Table~\ref{table_sel}) for the SCs, we chose to
select nights with 2 or more SCs present, rather than the 3 or
more for PCs.  With only 2 SCs, the NN approach is susceptible to
being less reliable than for the case of 3 PCs.  A wrong starting
value for one of 2 SCs would affect the computed transfer function
and, if the SC was only observed on one night (or a very few) with
only one other calibrator paired with it, an erroneous final value
could result. With 3 PCs, such a situation would have been
automatically corrected and, furthermore, 
the PCs were observed systematically on
more nights than the SCs.  

A run of the GS approach gave results for the SCs that initially
looked convincing but, when it was run again with initial
diameters changed by random amounts, it was found that the
solutions differed from those of the first run by more than the
formal errors on the diameters.  Therefore, in the case of the
SCs, the robustness of the solution needed further investigation.
A large number of simulations, in a MonteCarlo-style (MC) fashion,
were undertaken. For this, about 350 loops  of
the GS were executed. At each loop, the initial diameter guesses
were changed by random amounts from the nominal values.
The range of changes was set arbitrarily to $\pm7$\% of the
diameters. Each loop produced vectors of angular diameter and
error, one for each SC. From all the loops, the averages and
standard deviations of the angular diameters were deduced based on
the hypothesis of a Gaussian distribution of the results.
Note that the MC errors were dominant with respect to
the formal errors from a typical single GS run.

\begin{figure}
%\vspace{8cm}
\resizebox{\hsize}{!}{\includegraphics{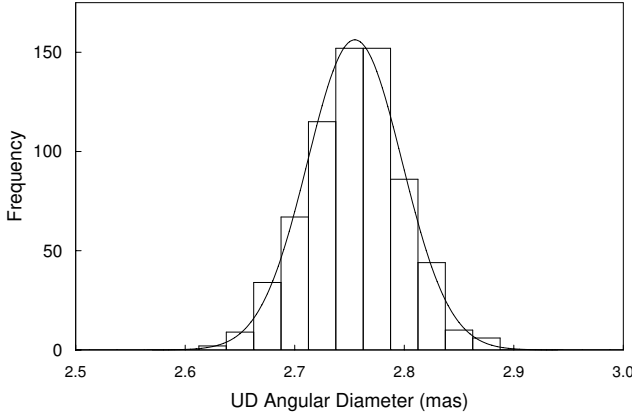}}
\caption[ ]{Result of the second and final MC
analysis for one of our secondary calibrators,
$\theta$ Cet.
The histogram represents the frequency of
solutions obtained from 739 loops of the GS approach, each with
random and independent starting values.
The line is a Gaussian fit, with average  and standard
deviation 2.755\,mas and 0.043\,mas, respectively.
}
\label{fig:mc1}
\end{figure}

\begin{table}
%\centering
\caption{List of the
angular diameters for the
secondary calibrators.
}
\label{tab:sc}
\begin{tabular}{crcr}
\hline
\hline
Star & \multicolumn{1}{c}{$\theta_{\mathrm{UD}}$(mas)} &
$\rho_{\lambda}$ & \multicolumn{1}{c}{$\theta_{\mathrm{LD}}$(mas)} \\
\hline
     HR 37 & $2.517\pm0.048$ &     1.0270 & $2.585\pm0.049$ \\

    20 Cet & $3.361\pm0.006$ &     1.0289 & $3.459\pm0.006$ \\

$\theta$ Cet & $2.755\pm0.043$ &     1.0225 & $2.817\pm0.044$ \\

$\delta$ Phe & $2.208\pm0.081$ &     1.0229 & $2.259\pm0.083$ \\

$\tau$ Cet & $2.030\pm0.084$ &     1.0200 & $2.071\pm0.086$ \\

$\chi$ Phe & $2.754\pm0.089$ &     1.0276 & $2.830\pm0.092$ \\

$\epsilon$ Eri & $2.081\pm0.064$ &     1.0220 & $2.127\pm0.065$ \\

$\delta$ Eri & $2.307\pm0.085$ &     1.0225 & $2.359\pm0.087$ \\

  39 Eri A & $1.677\pm0.085$ &     1.0265 & $1.722\pm0.087$ \\

$\gamma_1$ Cae & $2.173\pm0.071$ &     1.0261 & $2.230\pm0.073$ \\

$\eta_2$ Pic & $2.700\pm0.061$ &     1.0276 & $2.774\pm0.063$ \\

$\beta$ Ori & $2.881\pm0.177$ &     1.0114 & $2.914\pm0.179$ \\

   HR 1799 & $2.111\pm0.087$ &     1.0276 & $2.169\pm0.090$ \\

$\beta$ Lep & $2.942\pm0.064$ &     1.0208 & $3.003\pm0.066$ \\

    31 Ori & $3.495\pm0.030$ &     1.0272 & $3.590\pm0.031$ \\

   HR 2311 & $2.509\pm0.065$ &     1.0276 & $2.579\pm0.067$ \\

    30 Gem & $2.236\pm0.051$ &     1.0235 & $2.289\pm0.052$ \\

    17 Mon & $2.491\pm0.055$ &     1.0270 & $2.559\pm0.057$ \\

$\theta$ CMa & $4.032\pm0.198$ &     1.0265 & $4.139\pm0.204$ \\

$\delta$ CMa & $3.583\pm0.139$ &     1.0180 & $3.647\pm0.141$ \\

$\gamma_2$ Vol & $2.460\pm0.066$ &     1.0229 & $2.516\pm0.067$ \\

   HR 3046 & $1.726\pm0.106$ &     1.0229 & $1.765\pm0.108$ \\

$\iota$ Hya & $3.365\pm0.041$ &     1.0255 & $3.451\pm0.042$ \\

    31 Leo & $3.259\pm0.039$ &     1.0270 & $3.347\pm0.040$ \\

 $\xi$ Hya & $2.362\pm0.039$ &     1.0224 & $2.415\pm0.040$ \\

  V918 Cen & $3.000\pm0.029$ &     1.0276 & $3.082\pm0.030$ \\

   HR 4831 & $1.782\pm0.048$ &     1.0229 & $1.823\pm0.049$ \\

 $\pi$ Hya & $3.755\pm0.028$ &     1.0249 & $3.849\pm0.028$ \\

$\alpha$ Boo & $20.453\pm0.003$ &     1.0230 & $20.924\pm0.003$ \\

    51 Hya & $2.220\pm0.038$ &     1.0276 & $2.281\pm0.039$ \\

 $\mu$ Vir & $1.255\pm0.006$ &     1.0139 & $1.273\pm0.006$ \\

   HR 5513 & $2.136\pm0.016$ &     1.0273 & $2.194\pm0.016$ \\

    58 Hya & $3.297\pm0.046$ &     1.0260 & $3.383\pm0.048$ \\

$\phi_1$ Lup & $5.609\pm0.215$ &     1.0276 & $5.764\pm0.221$ \\

$\eta$ Ara & $5.515\pm0.179$ &     1.0276 & $5.667\pm0.184$ \\

   HR 6630 & $4.145\pm0.054$ &     1.0249 & $4.248\pm0.056$ \\

   HR 6862 & $2.597\pm0.037$ &     1.0273 & $2.668\pm0.038$ \\

$\lambda$ Sgr & $4.176\pm0.044$ &     1.0235 & $4.274\pm0.045$ \\

$\tau$ Sgr & $3.987\pm0.026$ &     1.0235 & $4.081\pm0.027$ \\

$\alpha$ Aql & $3.300\pm0.227$ &     1.0139 & $3.346\pm0.230$ \\

    56 Aql & $2.692\pm0.028$ &     1.0276 & $2.767\pm0.029$ \\

$\gamma$ Sge & $6.010\pm0.047$ &     1.0289 & $6.184\pm0.048$ \\

$\alpha$ Ind & $3.413\pm0.105$ &     1.0229 & $3.491\pm0.108$ \\

    24 Cap & $4.281\pm0.007$ &     1.0282 & $4.401\pm0.008$ \\

$\tau$ Aqr & $4.874\pm0.132$ &     1.0276 & $5.008\pm0.136$ \\

   HR 8685 & $1.957\pm0.071$ &     1.0289 & $2.013\pm0.073$ \\

    88 Aqr & $3.240\pm0.057$ &     1.0235 & $3.316\pm0.059$ \\
\hline \hline
\end{tabular}
\end{table}

After this first run, it was realized that for a few sources the
fixed range of $\pm7$\% was not sufficient to sample the
wings of the diameter distribution, and a second run (using
the results of the first run as input) was performed. For the
second run it was decided to adopt a range of random input values
between $\pm3$ and $\pm10$ times the formal accuracy of the
diameter as derived from the first run, depending on the star. A
total of about 750 loops were performed. 
An example of the distribution of resulting
diameters from this MC run for one particular star, $\theta$~Cet,
can be seen in Fig.~\ref{fig:mc1}.
We then adopted the mean and standard deviation of the fitting
Gaussian profiles for the diameters and their errors, and these
are listed
in Table~\ref{tab:sc}. 
Although the quality of the results may vary as reflected
by the errors, we note that all the MC distributions were
well fitted by Gaussian distributions.

These simulations only address the convergence of our numerical method; they say
nothing about the underlying assumptions and little, if anything about correlations
between the derived diameters and TF values. Both could result in errors larger than
those quoted here. Note that a global fit to the angular diameters and TF values fails
when all baselines are North-South (Mozurkewich, Per. Comm.). For each star, the
projected baseline is nearly constant through the night and therefore a decrease of all
the TF values cannot be differentiated from an increase of all the stellar diameters.
For some stars with an East-West baseline, the projected baseline length can change by
nearly a factor of two. This breaks the degeneracy between TF and the angular diameter
with the diameter determined from the fractional change of 
$|{\rm V}|^2$
through the night while
the determination of TF is from its mean value. The problem with East-West baselines
is that the projected baseline length correlates with zenith angle. Since the most likely
systematic variation of TF is with zenith angle (through a correlation of seeing with
zenith angle), East-West baselines are probably more susceptible to systematic errors
in the results. Including skewed baselines breaks this final degeneracy. Therefore it is
reasonable to expect that with measurements over a sufficient range of baseline lengths
and orientations, this algorithm of fitting for both the angular diameters and TF values
should work. As such it will help provide the long-sought solution to the problem of
calibrating OLBI data on very long baselines. However, these are still assumptions; the
sufficiently detailed simulation needed to address these issues is beyond the scope of this
paper.

\section{Results and discussion}\label{results}
\begin{figure}
%\vspace{5cm}
\resizebox{\hsize}{!}{\includegraphics{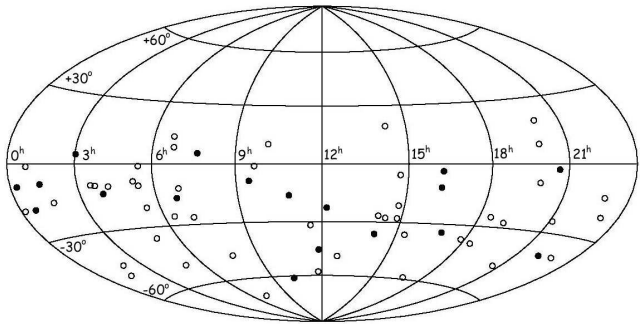}}
\caption[ ]{
Sky distribution of our primary (solid dots) and secondary
(circles) calibrators.
}
\label{fig:skycover}
\end{figure}

The sky coverage of our calibrators is relatively uniform
over the range $-80\degr$ to $+20\degr$, as shown in 
Fig.~\ref{fig:skycover}.
The average distance from any point in the sky, within the
above limits in declination i.e. within $45\degr$
zenithal distance at Paranal, to one of our
calibrators is $11\degr$. Specifically, the
average distance to a primary or
secondary calibrator is 20$\degr$ and $13\degr$, respectively.

The angular diameters of the primary calibrators listed in
Table~\ref{tab:gs} have an average accuracy of 0.35\%, with the
worst case being 0.90\%. This constitutes a remarkable achievement
for results based on direct observations and internal consistency
only. No assumptions were made for any angular diameters except
for estimated starting values for the night-by-night approach. In
any case, the resulting angular diameters are relatively
insensitive to the initial values.  A potential systematic effect
is associated with the values for the effective wavelengths used
in the fitting program.  The relative accuracy of 0.35\% implies a
knowledge of the effective wavelength for each star to better than
0.0077\,$\mu$m and we note that the effective wavelengths were
calculated to 0.001\,$\mu$m with an uncertainty of
$\pm$0.005\,$\mu$m \citep{Davis03}. An
additional uncertainty associated with the limb-darkened angular
diameters is the limb-darkening factor discussed in
Section~\ref{calib} but, as pointed out there, the adopted
values for $T_{\mathrm{e}}$, $\log g$, [Fe/H] and $\lambda$ for
each star affect the values for the factors at the level of
$<$0.1\%. Any uncertainty associated with the model atmosphere
predicted centre-to-limb intensity variations is unknown but is
likely to be at a comparable level. The accuracy of the primary
calibrator results presented here is superior to most single
determinations of angular diameters in published interferometric
results and therefore offers significant improvement over what is
currently available.

Many of our primary calibrators, being relatively bright, have had
values for their angular diameters published either from direct
measurements or from indirect estimates.  This provides us with
the opportunity for a comparison and a check for possible systematic
differences. As a first step, we have used the catalogue of
Cohen et al.
(\citeyear{Coh99}, CC hereafter) which lists limb-darkened
angular diameters based on absolutely calibrated stellar spectra,
and which has 12 stars in common with our list of primary
calibrators. The catalogue of calibrator stars for long baseline
stellar interferometry by Bord{\'e} et al. (\citeyear{Bor02}, BC
hereafter) includes the same 12 stars but adopts the Cohen et al.
values for the limb-darkened angular diameters and applies
limb-darkening corrections to give values for the uniform-disk
diameters in a number of spectral bands.  The comparison is
therefore made with the CC limb-darkened values.  The results are
listed in Table~\ref{tab:cohen} which lists our values, the CC
values, the differences between them in mas,  and the differences
in units of standard deviations (the latter taken as the sum of
the errors from our result and those quoted in CC).
%\begin{figure}
%%\vspace{5cm}
%\resizebox{\hsize}{!}{\includegraphics{richichi_fig7.eps}}
%%\includegraphics[width=84mm]{richichi_fig7.eps}
%\caption[ ]{
%A comparison of our LD angular diameters for the primary
%calibrators, against those from the catalogue of 
%\citet{Coh99}. The dotted line is the 1:1 relationship, and the
%solid line is a regression fit. Error bars are included but are
%difficult to be appreciated on this scale. }\label{fig:cohen}
%\end{figure}

%It can be seen from Table~\ref{tab:cohen} that there is no
%systematic difference between CC and our results.
%A plot of our limb-darkened angular diameters
%against those of CC is shown in Fig.~\ref{fig:cohen}. 
We performed a regression fit 
of the form $y=ax+b$, where $x$ are our values and $y$ those
of CC.
%, has been made and is also shown in Fig.~\ref{fig:cohen}.
Perfect agreement would result in $a=1$, $b=0$, and a regression
coefficient $R^2=1$.  The fit gave $a=1.016\pm 0.014$,
$b=-0.076\pm0.077$ and $R^2=0.998$.
Since the coefficients of the fit are consistent with their
expectation values, we conclude that there is no systematic bias
and that, for the 12 stars in common, our results and those of CC
show very good agreement. 
On average, the relative accuracy of our results
is about 3 times better than those of CC. 
The fact that the average 
$|\Delta \sigma|$  is slightly more than unity could be
a possible indication of some underestimation of the errors
in either of the two samples. 
We mention that we have
performed a detailed comparison also with the UD diameters of BC.
Although less compelling because it involves an additional
computation from LD to UD performed by the authors, this
comparison was also entirely satisfactory. Note that, due to the
different range of magnitudes, there is no overlap with the
catalogue of \citet{Mer05}. Other cases of
overlap exist between four of the stars in our primary calibrators
list and theoretical predictions given in 
\citet{bell}, \citet{alonso}, and 
\citet{decin}. Again there are no systematic differences and, for
those cases in which an error in the estimate is quoted by the
authors, the difference with our results is less than the sum of
the errors (i.e., $\Delta\sigma<1$ in the sense of
Table~\ref{tab:cohen}).

\begin{table}
\caption{ Comparison of our primary calibrator limb-darkened
angular diameters with the values by 
\citet{Coh99}.
$\Delta\sigma$ expresses the difference in units of the
combined error.
}
\label{tab:cohen}
\begin{tabular}{lrrrrrr}
\hline
\hline
  Star &
  \multicolumn{2}{c}{VLTI (V)} &
  \multicolumn{2}{c}{Cohen (C)} &
  \multicolumn{2}{c}{(V - C)} \\
    &
  \multicolumn{1}{c}{$\theta_{\rm LD}$} &
  \multicolumn{1}{c}{$\sigma$} &
  \multicolumn{1}{c}{$\theta_{\rm LD}$} &
  \multicolumn{1}{c}{$\sigma$} &
  \multicolumn{1}{c}{$\Delta$} &
  \multicolumn{1}{c}{$\Delta\sigma$} \\
    &
  \multicolumn{1}{c}{(mas)} &
  \multicolumn{1}{c}{(mas)} &
  \multicolumn{1}{c}{(mas)} &
  \multicolumn{1}{c}{(mas)} &
  \multicolumn{1}{c}{(mas)} &  \\
  \hline

$\iota$ Cet   &   3.325  &  0.010  &  3.36 &  0.036  &  -0.035  &  -0.9 \\
$\beta$ Cet   &   5.329  &  0.005  &  5.31 &  0.055  &   0.019  &   0.3 \\
$\gamma$ Eri  &   8.634  &  0.078  &  8.74 &  0.088  &  -0.106  &  -0.9 \\
  V337 Car    &   5.409  &  0.025  &  5.23 &  0.058  &   0.179  &   2.8 \\
$\nu$ Hya     &   4.652  &  0.015  &  4.57 &  0.048  &   0.082  &   1.6 \\
  HR 4546     &   2.538  &  0.004  &  2.60 &  0.040  &  -0.062  &  -1.5 \\
$\theta$ Cen  &   5.556  &  0.005  &  5.46 &  0.058  &   0.096  &   1.6 \\
$\chi$ Sco    &   2.176  &  0.013  &  2.10 &  0.023  &   0.076  &   2.9 \\
$\delta$ Oph  &   9.940  &  0.013  & 10.03 &  0.101  &  -0.090  &  -0.9 \\
$\epsilon$ Sco &  5.747  &  0.008  &  5.99 &  0.061  &  -0.243  &  -3.9 \\
  70 Aql      &   3.337  &  0.014  &  3.27 &  0.037  &   0.067  &   1.7 \\
$\lambda$ Gru &   2.703  &  0.020  &  2.71 &  0.030  &  -0.007  &  -0.2 \\
              &          &         &       &         &          &       \\
Average       &          &         &       &         &  -0.002  &   0.2 \\
\hline \hline
\end{tabular}
\end{table}

%\textbf{This next section needs rewriting since there are many
%more measurements than are currently listed in
%Table~\ref{tab:direct} and also 2 more stars (6 in total).}
%
A few direct measurements at 2.2\,$\mu$m have been published for
four of our primary calibrator stars, and a comparison with our
results is provided in Table~\ref{tab:direct}. Again, there is no
systematic discrepancy and the differences, in terms of standard
deviations as previously defined for Table~\ref{tab:cohen}, are
generally small. The only possible marginal agreement
is with the result by
Kervella et al. (\citeyear{Ker03}, K03 hereafter) for {$\alpha$
CMa}, but even in this case the disagreement is only
1.3 times the combined formal error, when the same quantity
is considered, namely the LD diameter.
%. We note that our measured quantity is the UD diameter, while
%K03 have fitted directly the LD diameter. If we compare this
%latter with our derivation of the LD value, there is in fact
%{\bf almost} an agreement (within {\bf 1.3} times the combined formal error).
We note that K03 used only one calibrator for their entire set of
long-baseline data namely {$\theta$ Cen}, for which they
used the diameter estimate from CC.  
%While
By comparison,
our result 
%is not necessarily better than that of K03, it should
%appears
%be more robust since it 
is based on nights with at least two
primary calibrators other than Sirius, and it is not based on any
assumptions of angular diameters.  Interestingly, the same authors
as K03 have obtained an angular diameter for {$\alpha$ CMi
A}, using VINCI and adopting {$\alpha$ CMa} as the only
calibrator, which is in excellent agreement with our value from
Table~\ref{tab:gs}.

\begin{table}
%\centering
\caption{
Comparison of our primary calibrator diameters with
available direct determinations.
}
\label{tab:direct}
\begin{tabular}{lrrrrr}
\hline
\hline
  \multicolumn{1}{c}{Star} &
  \multicolumn{2}{c}{UD} &
  \multicolumn{2}{c}{LD} &
  \multicolumn{1}{c}{Ref.} \\
    &
  \multicolumn{1}{c}{$\Delta$mas} &
  \multicolumn{1}{c}{$\Delta \sigma$} &
  \multicolumn{1}{c}{$\Delta$mas} &
  \multicolumn{1}{c}{$\Delta \sigma$} &
  \\
\hline
$\alpha$ Cet &      0.616 &        1.3 &            &            &          1 \\

%$\alpha$ CMa &      0.094 &    {\bf2.6}&      0.050 &   {\bf 1.3}&          2 \\
$\alpha$ CMa &      0.094 &    {2.6}&      0.050 &   { 1.3}&          2 \\

$\alpha$ CMi A &     -0.019 &       -0.3 &     -0.010 &       -0.1 &          3 \\

$\delta$ Oph &     -0.417 &       -0.8 &            &            &          4 \\

           &      0.363 &        0.7 &            &            &          5 \\

           &      0.363 &        0.9 &            &            &          1 \\

\hline
\hline
\end{tabular}
\\
References: 1) Dyck et al. (\citeyear{dyck06})
2) Kervella et al. (\citeyear{Ker03})
3) Kervella et al. (\citeyear{Ker04a})
4) Perrin et al. (\citeyear{perrin})
5) Dyck et al. (\citeyear{dyck08})
\end{table}

Several other direct measurements are available for some of our
primary calibrators. These include {$\alpha$ Cet}
(Mozurkewich et al. \citeyear{Moz03}); {$\gamma$ Eri}
(Mozurkewich et al. \citeyear{Moz03}); {$\alpha$ CMa} (Hanbury
Brown et al. \citeyear{hanbury}, Davis \& Tango \citeyear{davis86},
Mozurkewich et al. \citeyear{Moz03}); {$\alpha$ CMi A} (Hanbury
Brown et al. \citeyear{hanbury}, Mozurkewich et al. \citeyear{Moz03},
Nordgren et al. \citeyear{nordgren}); {$\alpha$ Hya}
(Mozurkewich et al. \citeyear{Moz03}); {$\delta$ Oph}
(Mozurkewich et al. \citeyear{Moz03}), as well as SUSI unpublished
measurements for {$\alpha$ CMa}, {V337 Car} and
{$\theta$ Cen}. However these measurements were obtained at
a range of wavelengths in the visible and details of bandpasses
and effective wavelengths are not generally available.  For this
reason, and since the limb-darkening corrections are larger and
possibly more uncertain for visible wavelengths than for
2.2\,$\mu$m, we have not made a detailed comparison of derived
limb-darkened angular diameters for these results.

\begin{figure}
%\vspace{5cm}
\resizebox{\hsize}{!}{\includegraphics{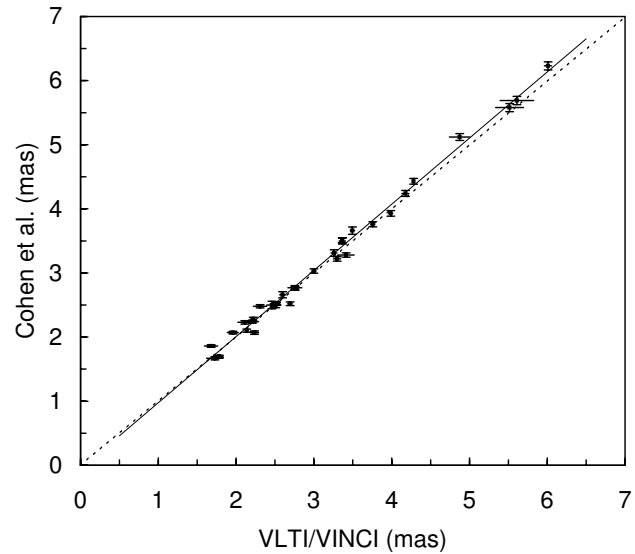}}
\caption[ ]{
A comparison of our LD angular diameters for the secondary
calibrators, against those from the catalogue of Cohen et al.
(\citeyear{Coh99}).
The dotted line is the 1:1 relationship, and the
solid line is a regression fit. Error bars are included but are
difficult to be appreciated on this scale.
%The same conventions of Fig.~\ref{fig:cohen} apply
%for the error bars and lines.
}
\label{fig:cohen-sc}
\end{figure}

Coming to the secondary calibrators, these have lower accuracy on
average than the primary ones,
as is to be expected from their more relaxed selection
criteria.
The average accuracy is 2.4\%, with
ten stars having accuracies better than 1\% and five stars between
5\% and 7\%. Not surprisingly, the smaller stars tend to have the
larger uncertainties, although the trend is not very strong and
cases of good accuracy can be found throughout the whole range of
angular diameters. Almost half of the stars have diameter errors
less than 0.05\,mas. We have performed the same comparison against
CC, as already done for the PCs. Thirty-two of our SCs are in
common with CC. The average difference is 0.4\,$\sigma$. A
regression fit gives values of $a=1.032\pm 0.016$,
$b=-0.057\pm0.053$ and $R^2=0.993$.
Fig.~\ref{fig:cohen-sc} illustrates the comparison.  Although the
agreement is not quite as good as for the PCs it can be regarded
nevertheless as satisfactory.

% The agreement is less ideal than for the PCs, but still rather
% convincing. Due to the relatively large number of stars in our SC
% sample, we have not attempted a comparison with direct
% measurements previously reported in the literature.

One could endeavour to find possible correlations of the
distribution of the differences between our results
and those of CC, with parameters such as magnitudes
or spectral types. However, our sample is rather
homogeneous in these respects. About 70\% of both
primary and secondary calibrators have early K
spectral types, and the remainder do not offer
sufficient range or statistics to
provide significant insight. The same applies
for the brightness, which for both types of
calibrators are concentrated mostly within 2 magnitudes.

%\section{Outline of future developments}\label{future}
%{\bf Probably this section can be abbreviated
%and become part of the conclusions}
%
%-extensions to other wavelengths
%
%-extensions to other baselines
%
%-extensions to other instrument modes
%
%-tools, WEB
%
%-schedule
%
%-collaborations

\section{Conclusions}
We have undertaken what is probably 
%one of the first and largest
a unique
attempt to derive a homogeneous list of interferometric
calibrators, based entirely on observations with the
same instrument and without influence
from pre-existing estimates. 
The stars in our list cover a range
of magnitudes from $\approx$$-3$ to $\approx$$+3$ in the K-band,
and from 1.3 to 20\,mas in diameter. Roughly 70\% of the stars
have diameters in the 2-5\,mas range.
Similarly to work done in the area of photometric systems,
we have divided our sample into two groups of 17 primary and 47 secondary
calibrators. The former have a broader and more detailed database
of observations, which has led to a high accuracy and robustness of
the results. The primary calibrators have then been used as
reference points to help build up the secondary calibrators
using a numerical approach to obtain a global solution.

Our calibrators are appropriate for
observations in the near-infrared with baselines in
the range 20-200\,m obtained with aperture sizes in the
1\,m class, and as such it can satisfy the needs of most
current interferometers. 
It is clear that, for observations with
very long baselines and very large apertures (such as the 8.2\,m
telescopes of the VLTI), a new approach to interferometric
calibration will be needed.  This will be necessarily
based more on computed
diameters than on measured ones.  In this case, our calibrators are well
suited to provide the crucial first step of validating theoretical
predictions against measured values.  
This is especially true in
the case of the primary calibrators, with accuracies  that provide
an exacting test (0.35\% on average, 0.9\% in the worst case) for
theoretical predictions.
A comparison with the widely used catalogue of indirect estimates
of angular diameters by \citet{Coh99}, as well as with other similar
catalogue and with previous direct measurements when available,
gives in general very good agreement. We note however that, at
least for the primary 
calibrators, the internal accuracy achieved in the present work
is on average three times better than that of
\citet{Coh99}.

There are some limitations in the present work that must be born
in mind. One approximation has been to assume that the transfer
function remained constant throughout each night, i.e. one unique
value for each night.  We have given practical and empirical
considerations to justify this choice.
The complications involved in the inclusion
of variations of the transfer function during the course of each
night were beyond the scope of the present study.  However,
refinements in this approach
are indeed important and
should be considered in future work wherever feasible.

% Mathematical complications, as well as the need to ensure a
% significative number of nights and calibrators for our analysis,
% have adviced us to neglect this point in the present work. Future
% similar efforts should try to include variations of the transfer
% function.

Another important issue is the correction for limb darkening,
which is essential if one wants to relate the angular diameter
observed at one wavelength to the work done at another wavelength,
or if the diameter is to be used as a parameter in comparisons of
observations with stellar atmospheric models.  The corrections we have
applied are based on model atmosphere predictions of
centre-to-limb variations of intensity and they are therefore only
as good as the models.  However, if improved model predictions
become available, our uniform-disk angular diameters remain
unchanged and the limb-darkened diameters are readily updated.

% We have discussed and computed limb-darkening factors for the
% stars in our sample and we are confident that, to the level of
% accuracy required, the corrections we have applied are consistent
% with the currently available models.

% However, it is worth stressing that, especially for high accuracy
% measurements of some extreme spectral types, the uncertainty in
% limb-darkening corrections can become significant. It is important
% to improve our knowledge in this area. This cannot be done through
% theory alone, but requires dedicated observations of challenging
% difficulty.

%\begin{acknowledgements}
\section*{Acknowledgments}
This research is
based on observations made with the European Southern Observatory
telescopes obtained from the ESO/ST-ECF Science Archive Facility,
and has made use of the \textit{Simbad} database, operated at the
Centre de Donn{\'e}es Astronomiques de Strasbourg (CDS), and of
NASA's Astrophysics Data System Bibliographic Services (ADS).
We acknowledge the detailed comments by the
the referee Dr. D. Mozurkevich, who has contributed the last
paragraph of Sect.~\ref{analysis_sec}.
AR's visits to the School of Physics at the University of
Sydney in 2004 and 2006 were supported by an ESO
Director General Discretionary Fund grant.
JD was a visiting scientist at the ESO Headquarters
in Garching during April and May 2005.
This work is entirely based on the
vast volume of data obtained with the
VINCI instrument and made public 
in the first years of VLTI commissioning.
This was made possible through the effort and
dedication of a large group of ESO astronomers, engineers
and students.
%\end{acknowledgements}

\appendix
\section{Wavelength and bandpass issues}\label{app}
In developing a list of interferometric calibrators we are
concerned with accuracies in the angular diameters of the order of
a few parts in a thousand.
At this level, the precise definition of angular diameter
as a function of wavelength 
becomes essential, and 
in particular it is necessary to account for
limb darkening.  Similarly,
it is crucial to understand in detail
the effect of the filter bandpass.
\subsection{Limb Darkening corrections}
The observed
visibility values should properly be fitted with the transform for
a limb-darkened disk with the centre-to-limb brightness
distribution as an additional unknown.  However, for stars with
compact atmospheres \citep{Bas91}, i.e. stars for
which the thickness of their atmosphere is very small compared to
their radius, the differences between the \textit{shape} of the
transform for a uniformly illuminated disk and that for a
limb-darkened disk in the region shortward of the first zero in
the transform are too small to be measured with sufficient
accuracy to distinguish between them reliably. The \textit{scale}
of the transform for a limb-darkened disk will differ from that
for a uniform-disk depending on the wavelength of observation and
on the effective temperature, surface gravity and metallicity of
the star.  The generally adopted approach is to fit the transform
for a uniform-disk to the measurements of visibility squared to
obtain the equivalent uniform-disk angular diameter,
$\theta_{\mathrm{UD}}$, and to apply a correction determined from
model stellar atmospheres to obtain a value for the true
limb-darkened angular diameter, $\theta_{\mathrm{LD}}$. The
transform for a uniform-disk is

\begin{equation}
V^{2} = \left|\frac{2J_{1}(\pi
b\theta_{\mathrm{UD}})}{\lambda_{\mathrm{eff}}}\right|^{2}
\label{eq:transform}
\end{equation}
where $J_{1}$ is a Bessel function, $b$ is the projected baseline
and $\lambda_{\mathrm{eff}}$ the effective wavelength of
observation.  Eqn.~\ref{eq:transform} has been fitted to the
observational data to determine values for $\theta_{\mathrm{UD}}$.
The uniform-disk angular diameters are only applicable for the
effective wavelength at which they were measured.  For values at
other wavelengths the effects of limb-darkening must be taken into
account.  This generally involves scaling the measured
uniform-disk angular diameter to the equivalent limb-darkened
angular diameter and then scaling it to the uniform-disk angular
diameter for the desired wavelength.  Since the selected primary
calibrators are single, non-rotating or slowly rotating stars with
compact atmospheres, the limb-darkening scaling factors can be
determined from grids of theoretical model stellar atmospheres.
\citet{Davis00} have computed limb-darkening factors
$\rho_{\lambda}$ as a function of effective temperature
($T_{\mathrm{e}}$), surface gravity ($\log g$), metallicity
([Fe/H]) and wavelength ($\lambda$) for the entire grid of model
atmospheres by \citet{Kur93a, Kur93b}.
%, \citealt{Kur93b}).
\citet{Claret00} has computed non-linear limb-darkening
coefficients for the same model atmospheres and these lead to the
same results for the limb-darkening factors. With
$\theta_{\mathrm{UD}}$ the uniform-disk angular diameter at
wavelength $\lambda$ and $\theta_{\mathrm{LD}}$ the corresponding
limb-darkened angular diameter, the limb-darkening factor
$\rho_{\lambda}$ is defined by

\begin{equation}
\label{eq:ld} \rho_{\lambda} =
\frac{\theta_{\mathrm{LD}}}{\theta_\mathrm{UD}}
\end{equation}

In order to obtain the limb-darkening factors we have interpolated
within the tabulations given by \citet{Davis00} with
values for $T_{\mathrm{e}}$, $\log g$, [Fe/H] and $\lambda$ for
each star. For $\lambda$, the value of the effective wavelength of
the measurement ($\lambda_{\mathrm{eff}}$), as discussed in
Section~\ref{pri_sec}, was adopted. Values for $T_{\mathrm{e}}$,
$\log g$ and [Fe/H] have been obtained for most of the calibrators
from the tabulations of \citet{Str97, Str01}. 
For the remaining calibrators, values have been adopted
by inspection and interpolation among the values for the other
stars. We note that the uncertainties in the adopted values only
affect the limb-darkening factors at the level of 0.1\%.  The
values for $\lambda_{\mathrm{eff}}$ and $\rho_{\lambda}$ for each
star are listed in the Tables~\ref{tab:nn} and \ref{tab:sc}. The
limb-darkened angular diameters that we list are, following
Eqn.~\ref{eq:ld}, the product of the uniform-disk angular
diameters and the limb-darkening factors.
\subsection{Bandwidth smearing}
The bandpass of VINCI is
1.92-2.50$\mu$\,m at the 1\% level 
\citep{Davis03}, and therefore each visibility point is the
average, according to proper weights of instrument sensitivity,
source spectral distribution and atmospheric transmission, 
of a wide range of monochromatic
visibilities. While this has almost no effect in the linear part
of the visibility curve, in the non-linear regime of very low
visibility values the broad spectral bandwidth might result in a
significant effect on the measured visibilities. There is the
further complication that the monochromatic uniform-disk angular
diameter will vary across the bandpass. We have performed
numerical investigations of this effect for a number of extreme
cases among our sample, in particular for the three primary
calibrators {$\alpha$ CMa}, {$\epsilon$ Crv} and
{V337 Car}, and the secondary calibrator {V918 Cen}.
All these objects had some calibrated visibility squared values as
low as $\approx$~0.1 and were therefore expected to be most
sensitive to bandwidth smearing.  The differences that we found
between the diameter derived from a monochromatic analysis at the
assumed effective wavelength, and the diameter derived from a
detailed multi-wavelength analysis, were smaller than the formal
errors in the diameters. 
This effect is systematic but it is not necessarily in the same
direction as it depends on the baselines used.  For stars with only
short baseline observations the fitted uniform-disk angular
diameters will be too large but, for long baseline observations or
a range of baselines including long ones, the fitted diameters
will be too small.
Our
investigation showed that the effect of bandwidth smearing could
be significant if only low visibility values were used in the
angular diameter fitting program.  However, in general,
observations at several baselines were available for the fitting
program giving a combination of high and low visibilities. 
Only in the case of 
%{$\epsilon$ Crv} 
{$\alpha$ CMa}
was the difference comparable with the formal error.  
%This
%reduced the effect of bandwidth smearing on the angular diameter
%to an insignificant level except in the case of {$\epsilon$
%Crv} for which only low visibility observations were available.
Given these results it was decided to ignore, with considerable
simplification of the data analysis, the effect of bandwidth
smearing for all stars, and to increase the formal error by a
factor of two for 
{$\alpha$ CMa} 
alone.

\end{document}